\documentclass{article}
\usepackage{amsfonts}
\usepackage{amsmath}
\usepackage{amsthm}
\usepackage{epsfig}

\newcommand{\dt}{\frac{d}{dt}}
\newcommand{\drs}{\frac{\partial}{\partial r_*}}
\newcommand{\drss}{\frac{\partial^2}{\partial r_*^2}}
\newcommand{\Reals}{\mathbf R}

\newcommand{\defin}{\equiv}
\newcommand{\like}{\approx}
\newcommand{\starman}{\mathfrak{M}}

\newcommand{\hide}[1]{}
\newcommand{\horifac}{(1-\frac{2M}{r})}

\newcommand{\ProvenInBS}[1]{ Proof: in #1 of \cite{BS}}

\renewenvironment{proof}{ Proof:}{$\Box$}
\newtheorem{theorem}{Theorem}[section]

\newtheorem{lemma}[theorem]{Lemma}

\newtheorem{definition}[theorem]{Definition}

\title{The wave equation on the Schwarzschild metric\\ II: Local Decay for the spin 2 Regge Wheeler equation}
\author{P. Blue\\ \small Mathematics dept., Rutgers University \\ \small Piscataway, NJ 08854, USA\\ \and A. Soffer\\ \small Mathematics dept., Institute for Advanced Studies\\ \small Princeton, NJ, 08540, USA}

\begin{document}
\maketitle

Recently, it has been shown that the wave equation for a scalar field on the exterior part of the Schwarzschild manifold satisfies local decay estimates useful for scattering theory and global existence \cite{BS}. The extension for the linearized Einstein equation is considered here. In 1957, Regge and Wheeler investigated spin 2 tensor fields on the Schwarzschild manifold \cite{RW}. They classified such fields into two types, which they called even and odd. For the odd fields, they were able to reduce the problem to an equation for a scalar field very similar to the wave equation for scalar fields on the Schwarzschild manifold. In 1970, Zerilli extended their results to include the even case; although, the equation for the even case is significantly more complicated and bears less resemblance to the wave equation for a scalar field \cite{Z}. Teukolsky has done a related reduction for the rotating Kerr black hole \cite{Teukolsky} which has been used to investigate the stability of the black holes \cite{Whiting}. 

This paper extends the local decay estimate for the scalar wave equation of \cite{BS} to the Regge-Wheeler equation. Many of the proofs used here follow \cite{BS}. We obtain the following for $r_*$ the standard Regge-Wheeler co-ordinate and $\beta>\frac{3}{2}$, there is a constant $C$, depending on the initial condition through the energy norm, so that
\begin{equation*}
\int_0^\infty \|(1+(\frac{r_*}{2M})^2)^{-\frac{\beta}{2}}u\|^2dt <C
\end{equation*}

\section{Co-ordinates and equations}
The Schwarzschild manifold describes a static black hole solution to the Einstein equation. The exterior of the black hole is most easily described by $(t,r,\theta,\phi)\in\Reals\times(2M,\infty)\times S^2$ with the metric
\begin{equation}
ds^2=\horifac dt^2-\horifac^{-1} dr^2-r^2 ds_{S^2}^2
\end{equation}

To simplify the problem, Regge and Wheeler (\cite{RW}) introduced a new radial co-ordinate, $r_*$, satisfying
\begin{equation}
\label{eDiffDefRS}
\frac{\partial r}{\partial r_*}=\horifac
\end{equation}
This allows the definition of a space like manifold
\begin{equation}
\starman=\Reals\times S^2
\end{equation}
The old co-ordinate $r$ is now treated as a function of $r_*$. 

In these new co-ordinates, the Regge-Wheeler equation for a scalar field $u:\Reals\times\starman\rightarrow\Reals$ which determines the behavior of the odd-type tensor fields is
\begin{equation}
\label{RWeqn}
\ddot{u}+Hu=0
\end{equation}
where
\begin{align}
H=&\sum_{j=1}^3 H_j \\
H_1=& -\drss \\
H_2=& (1-s^2)V \\
V=& \frac{2M}{r^3}\horifac\\
H_3=& V_L (-\Delta_{S^2}) = V_L\sum_{l=0}^\infty l(l+1)P_l\\
V_L=& \frac{1}{r^2}\horifac
\end{align}
and where $s=2$ for the case of the tensor field and $P_l$ is projection onto spherical harmonics with total angular momentum $l$. The case $s=0$ is the scalar field previously considered and $s=1$ is for the odd-type vector (Maxwell) fields.  

Because of the way the scalar field $u$ is defined it is not possible for it to have any component with spherical harmonic component $l=0$. It has also been shown that the $l=1$ component corresponds to changing the non-rotating Schwarzschild background to a rotating Kerr solution and to gauge transformations \cite{SarbachTiglio}. For this reason, we only consider $u$ with no $l=0$ or $l=1$ spherical harmonic component. This provides a lower bound on the spherical Laplace-Beltrami operator. 
\begin{equation}
\label{minLcondition}
-\Delta_{S^2}\geq2(2+1)=6
\end{equation}

For the scalar wave equation, Bachelot and Nicolas have proven global existence \cite{BachelotNicolas} in both an energy space and in $C^{\infty}$. The assumption of global existence in $C^\infty$ greatly simplifies all the following arguments and will be assumed although we are not yet aware of a published proof. However, the method of Bachelot and Nicolas should extend to the Regge-Wheeler equation without difficulty. The assumption of global existence in $C^\infty$ means that all solutions are assumed to be $C^{\infty}(\starman)\cap H^1(\starman,dr_*d^2\omega_{S^2})$, are infinitely differentiable in $t$ and have time derivative in $C^{\infty}(\starman)\cap L^2(\starman,dr_*d^2\omega_{S^2})$. The notation $u(t)$ denotes the function from $\starman\rightarrow\Reals$ corresponding to $u$ evaluated at time $t$. The measure $dr_*d^2\omega_{S^2}$ is used for all norms and inner products unless otherwise specified. 

\section{The Heisenberg-type relation and preliminary estimates}

For the Schr\" odinger equation, the Heisenberg relation describes the time evolution of expectation values for an operator and gives conserved quantities from symmetries of the Hamiltonian. A similar relation exists for the wave equation \cite{BS}. 

\begin{theorem}[Heisenberg-like relation]
\label{THeisenberglike}
For a time independent operator $A$ and a solution to the linear wave equation $\ddot{u}+Hu=0$ such that $u$ and $Hu$ are in the domain of $A$, and $u$ and $Au$ are in the domain of $H$,
\begin{equation}
\dt(\langle u,A\dot{u}\rangle-\langle\dot{u},Au\rangle)=\langle u,[H,A]u\rangle
\end{equation}
\end{theorem}
\ProvenInBS{Theorem 1}

The first and most important application of theorem \ref{THeisenberglike} is conservation of energy. As usual it is generated by time translation symmetry. This result is already well known \cite{Wald}. 

\begin{theorem}[Energy Conservation]
\label{TEnergyConservation}
The Regge-Wheeler equation, equation \ref{RWeqn}, has a conserved quantity $\|u\|_\mathcal{H}^2$ which we call the energy.
\begin{equation}
\|u\|_\mathcal{H}^2 = \langle\dot{u},\dot{u}\rangle+\langle u',u'\rangle+\langle u,-3Vu\rangle +\sum_{l=2}^\infty\langle P_l u,l(l+1)V_L P_l u\rangle
\end{equation}
This acts as a metric on the space $\mathcal{H}=\{u\in L^2: \|u\|_{\mathcal{H}}<\infty\}$. 
\end{theorem}
\begin{proof}
The conservation of energy follows from the Heisenberg-like relation with the multiplier $A=\dt$. This acts as a metric because the $\langle u,(H_2+H_3)u\rangle$ term is positive. The positivity of this potential was known to Regge and Wheeler \cite{RW} and is verified here. 

Since only functions orthogonal to the spherical harmonics with $l=0$ and $l=1$ are considered
\begin{align*}
\langle u,-3Vu\rangle +\sum_{l=2}^\infty\langle P_l u,l(l+1)V_l P_l u\rangle \geq&
\langle u,-3V u\rangle+\langle u,6V_l u\rangle\\
\geq&\langle u,(\frac{-6M}{r^3}+\frac{6}{r^2})\horifac u\rangle\\
\geq&\langle u, \frac{6}{r^2}(1-\frac{M}{r})\horifac u\rangle
\end{align*}
Since $\frac{6}{r^2}(1-\frac{M}{r})\horifac$ is always positive $\langle u,(H_2+H_3)u\rangle$ is positive definite. Since $\langle \dot{u},\dot{u}\rangle$, $\langle u',u'\rangle$, and $\langle u,(H_2+H_3)u\rangle$ are all positive each is defined if $\|u\|_{\mathcal{H}}$ is finite, and $\|u\|_\mathcal{H}$ is a metric. 

\end{proof}

As stated in the proof of energy conservation, the energy controls certain derivative norms and this can be used to control the growth of the $L^2$ norm. 

\begin{theorem}
If $u$ is a real valued solution of the Regge-Wheeler equation (equation \ref{RWeqn}) and $\|u(t)\|_{L^2}$ is the norm of $u$ at time $t$, then for $t\geq 0$
\begin{align*}
\|\dot{u}\|_{L^2}\leq&\|u\|_\mathcal{H}\\
\|u(t)\|\leq&t\|u\|_\mathcal{H} +\|u(0)\|_{L^2}
\end{align*}
\end{theorem}
\begin{proof}
Since $\langle u',u'\rangle$ and $\langle u,(H_2+H_3)u\rangle$ are strictly positive $\|\dot{u}(t)\|$ is controlled by the energy. This is used to control the growth rate of $\|u(t)\|_{L^2}$
\begin{align*}
\dt\|u(t)\|_{L^2}^2=&\dt\langle u,u\rangle =2\langle u,\dot{u}\rangle\\
2\|u\|_{L^2} \dt\|u\|_{L^2}\leq& 2\|u\|_{L^2} \|\dot{u}\|_{L^2}\\
\dt\|u\|_{L^2}\leq& \|\dot{u}\|_{L^2} \leq \|u\|_\mathcal{H}
\end{align*}
\end{proof}

\section{Local Decay}
For the scalar wave equation, a radial differential operator $\gamma$ was introduced to prove the local decay estimate \cite{BS}. It is used here for the same purpose. This multiplier is centered at the peak of the angular potential $V_L$. To simplify calculations the standard $r_*$ co-ordinate is translated to a new one, $\rho_*=r_*-\alpha_*$ so that $\rho_*=0$ at the peak of $V_L$. This centrifugal tortoise co-ordinate satisfies the same differential definition as $r_*$, equation \ref{eDiffDefRS}. 

\begin{definition}
The centrifugal origin $\alpha$, centrifugal tortoise origin $\alpha_*$, and centrifugal tortoise radius $\rho_*$ are defined by
\begin{align}
\alpha\defin&3M\\
\alpha_*\defin&r_*|_{r=\alpha=3M}\\
\rho_*\defin&r_*-\alpha_*
\end{align}
\end{definition}

\begin{definition}
Given $\sigma\in(\frac{1}{2},1]$, the Morawetz type multiplier $\gamma_\sigma$ is defined by
\begin{align}
g_\sigma(r_*)\defin&\int_0^{\frac{r_*-\alpha_*}{2M}}(1+\tau^2)^{-\sigma}d\tau\\
\gamma_\sigma\defin&-\frac{i}{2}(g_\sigma\drs +\drs g_{\sigma})
\end{align}
As before, $C^\infty$ solutions are assumed so that there are no domain issues. In all cases. the value of $\sigma$ will be fixed and the notation $g=g_\sigma$ and $\gamma=\gamma_\sigma$ will be used. 
\end{definition} 

\begin{theorem}
If $u\in H^1(\starman)$ and $\sigma\in(\frac{1}{2},1]$, then
\begin{equation}
\langle u,\gamma_\sigma u\rangle =0
\end{equation}
\label{5.1}
and there is a constant $C_\sigma=\lim_{r_*\rightarrow\infty}g(r_*)$ such that
\begin{equation}
\label{5.2}
\|\gamma u\|\leq C_\sigma\|u\|_\mathcal{H}+\frac{1}{2}\|(1+(\frac{r_*-\alpha_*}{2M})^2)^{-\sigma}u\|_{L^2}
\end{equation}
\end{theorem}
\begin{proof}
Equation \ref{5.1} is proven in theorem 16 of \cite{BS} (the statement of which includes the additional, but unnecessary, assumption that $u$ satisfy the scalar wave equation). 
For equation \ref{5.2}, theorem 17 of \cite{BS} does not directly apply since the space $\mathcal{H}$ defined there involves different potentials. However, the same argument applies. It is first noted that since $\sigma>\frac{1}{2}$ and the integrand in the definition of $g$ is positive and even, $|g|$ is bounded by $C_\sigma=\lim_{r_*\rightarrow\infty}g(r_*)$. Now, by direct computation,
\begin{align*}
\|\gamma u\|=&\|gu'+\frac{1}{2}g'u\|\\
\leq&\|gu'\|+\frac{1}{2}\|g' u\|\\
\leq& C_\sigma \|u\|_\mathcal{H}+\frac{1}{2}\|(1+(\frac{r_*-\alpha_*}{2M})^2)^{-\sigma}u\|_{L^2}
\end{align*}
\end{proof}

The Heisenberg-like relation will be applied to the multiplier $\gamma$. To do this it is necessary to estimate the commutator $[\sum_{j=1}^3H_j,\gamma]$. 

\begin{lemma}
For $\sigma\in(\frac{1}{2},1]$
\begin{equation}
\label{eH3lowerbound}
i[\frac{-1}{r^2}\horifac \Delta_{S^2}, \gamma]=-g(r_*)[\frac{3M}{r}-1]\frac{2}{r^3}\horifac\Delta_{S^2} \geq 0
\end{equation}
\end{lemma}
\ProvenInBS{Lemma 18}

\begin{lemma}
For $\sigma\in(\frac{1}{2},1]$ and $u$ in the domain of $\gamma$ and $H$
\begin{equation}
\label{eH1lowerbound}
\langle u, i[-\drss,\gamma]u\rangle \geq \langle u,\frac{\sigma}{(1+(\frac{\rho_*-\alpha_*}{2M})^2)^{\sigma+2}}\frac{1}{(2M)^3}[5+(3-2\sigma)(\frac{\rho_*-\alpha_*}{2M})^2]u\rangle
\end{equation}
\end{lemma}
\ProvenInBS{Lemma 20}

\begin{lemma}
\label{LMorawetzlike}
For $\sigma\in(\frac{1}{2},1]$ and $u$ in the domain of $H$ and $\gamma$, there is a constant $c_\sigma$ so that 
\begin{equation}
\label{6.1}
\langle u, i[\sum_{i=j}^3H_j,\gamma]\rangle\geq\langle u,\frac{c_\sigma}{(1+(\frac{r_*-\alpha_*}{2M})^2)^{\sigma+1}}u\rangle
\end{equation}
\end{lemma}
\begin{proof}
Since $\sigma\leq1$
\begin{equation*}
\frac{\sigma}{(1+(\frac{\rho_*-\alpha_*}{2M})^2)^{\sigma+2}}\frac{1}{(2M)^3}[5+(3-2\sigma)(\frac{\rho_*-\alpha_*}{2M})^2] < \frac{\sigma}{(1+(\frac{\rho_*}{2M})^2)^{\sigma+1}}
\end{equation*}
In the proof of lemma 21 of \cite{BS} it is shown that 
\begin{equation}
\label{eVlowerbound}
i[V,\gamma]= g(3-\frac{8M}{r})\frac{2M}{r^4}\horifac
\end{equation}
Since $g<0$ for $r<3M$ and $g>0$ for $r>3M$, $i[H_2,\gamma]=-3i[V,\gamma]$ is negative for $r<\frac{8M}{3}$, positive for $\frac{8M}{3}<r<3M$, and negative for $3M<r$. In the region $\frac{8M}{3}<r<3M$, all the terms of the form $i[H_j,\gamma]$ are positive so an estimate of the form \ref{6.1} holds. The other $r$ values are now treated. 

It is useful to note that a term relating $H_2$ to $H_3$ is decreasing since
\begin{equation*}
\frac{d}{dr} \frac{3-\frac{8M}{r}}{1-\frac{3M}{r}} =-\frac{M}{(r-3M)^2} <0
\end{equation*}

At $r=2M$, $3-\frac{8M}{r}=-1=2(1-\frac{3M}{r})$. Therefore in $2M<r\leq\frac{8M}{3}$ and for $l\geq2$
\begin{align*}
|3-\frac{8M}{r}|\leq&2|1-\frac{3M}{r}|\\
| r\frac{1}{r^3}\horifac(\frac{2M}{r})(3-\frac{8M}{r})|<&|g\frac{1}{r^3}\horifac 2(1-\frac{3M}{r})|\\
|i[H_2,\gamma]|<&\frac{1}{6}i[H_3,\gamma]
\end{align*}
The $\frac{1}{6}$ factor is present due to the restriction that $l\geq2$ and hence $-\Delta_{S^2}\geq6$. 

At $r=3.3M$
\begin{equation*}
\left| \frac{2(1-\frac{3M}{r})6}{-3(3-\frac{8M}{r})}\right|=\frac{1\frac{1}{11}}{1\frac{8}{11}}=\frac{12}{19}>\frac{2M}{3.3M}
\end{equation*}
Therefore for $r>3.3M$ and $l\geq2$
\begin{align*}
|-3(3-\frac{8M}{r})\frac{2M}{3.3M}|<&2(1-\frac{3M}{r})6\\
|\frac{g}{r^3}\horifac(-3)(3-\frac{8M}{r})\frac{2M}{r}|<&\frac{g}{r^3}\horifac(1-\frac{3M}{r})(-\Delta_{S^2})\\
|i[H_2,\gamma]|<&i[H_3,\gamma]
\end{align*}

Finally for $3M\leq r\leq3.3M$, since $i[H_3,\gamma]$ vanishes quadratically in $(r-3M)$ where as $i[H_2,\gamma]$ vanishes only linearly it is necessary to bound $i[H_2,\gamma]$ by $i[H_1,\gamma]$. On this interval $F^{-1}<3$, $\rho_*<.9M$, and $g<.9M$. Again, assuming $l\geq2$
\begin{align*}
|i[H_2,\gamma]|=&3g(3-\frac{8M}{r})(\frac{2M}{r})^4\frac{1}{(2M)^3}\horifac\\
&<(3)(.9)(\frac{19}{33})(\frac{2}{3})^4(\frac{13}{33})\frac{1}{(2M)^3}< .121\frac{1}{(2M)^3}\\
i[H_3,\gamma]>& \frac{5\sigma}{(1+(\frac{\rho_*}{2M})^2)^{\sigma+2}}\frac{1}{(2M)^3}>\frac{5\sigma}{(1.2025)^{\sigma+2}}\frac{1}{(2M)^3}>1.43\frac{1}{(2M)^3}
\end{align*}

In summary, for $r<3M$ and $r>3.3M$, $i[H_2+H_3,\gamma]>0$ and for $3M\leq r\leq3.3M$, $i[H_1+H_2,\gamma]$ is strictly positive.  Since for $r<3M$ and $r>3.3M$, $i[H,\gamma]>i[H_1,\gamma]>C(1+(\frac{\rho_*}{2M})^2)^{-\sigma-1}$ and for $3M\leq r\leq3.3M$, $i[H,\gamma]$ is strictly positive, there is a constant $C$ so that
\begin{equation*}
\langle u,i[H,\gamma]u\rangle \geq \langle u,\frac{C}{(1+(\frac{r_*-\alpha_*}{2M})^2)^{\sigma+1}}u\rangle
\end{equation*}
\end{proof}

It is now possible to apply the Heisenberg type relation to $\gamma$ and integrate the result to prove local decay. 

\begin{theorem}[Local Decay]
If $u$ is a solution to the Regge-Wheeler equation (equation \ref{RWeqn}), $\|u\|_\mathcal{H}=E$, $u(0)=f$, and $\beta>\frac{3}{2}$, then there is a constant $D_\sigma$ such that
\begin{equation}
\int_0^\infty \|(1+(\frac{r_*}{2M})^2)^{-\frac{\beta}{2}}u\|^2 dt\leq D_\sigma E^\frac{1}{2}(E^\frac{1}{2}+\|f\|_{L^2})
\end{equation}
\end{theorem}
\begin{proof}
Initially the result will be proven with $\beta=\sigma+1$ and $\sigma\in(\frac{1}{2},1]$ and $\rho_*$ in place of $r_*$. By integrating lemma \ref{LMorawetzlike} and applying the Heisenberg like relation, Theorem \ref{THeisenberglike}, it is possible to bound the time integral of the local decay term by an inner product evaluated at time $T$. Despite the explicit factors of $i$ appearing in the following, all terms are real valued. 
\begin{align}
&\int_0^\infty \|(1+(\frac{r_*-\alpha_*}{2M})^2)^{-\frac{\sigma+1}{2}}u\|^2 dt\nonumber\\
&\leq \int_0^T i[H,\gamma] dt\nonumber\\
&\leq \int_0^T \dt( \langle u, i\gamma\dot{u}\rangle-\langle\dot{u},i\gamma u\rangle)dt\nonumber\\
&\leq \int_0^T i\dt( \dt\langle u,\gamma u\rangle -2\langle\dot{u},\gamma u\rangle)dt \nonumber\\
&\leq \int_0^T -2i\dt\langle\dot{u},\gamma u\rangle dt\nonumber\\
&\leq 2(\|\dot{u}\| \|\gamma u\|)|_{t=T}+2(\|\dot{u}\| \|\gamma u\|)|_{t=0}\nonumber\\
&\leq E^{\frac{1}{2}}(4C_\sigma E^{\frac{1}{2}}+\|(1+(\frac{r_*-\alpha_*}{2M})^2)^{-\sigma} f\|) +E^{\frac{1}{2}}\|(1+(\frac{r_*-\alpha_*}{2M})^2)^{-\sigma} u(T)\|\label{lastlinebeforeHolder}
\end{align}
Since $\sigma>\frac{1}{2}$, $q$ can be chosen so that $\frac{1}{2\sigma}+\frac{1}{2}<q<\frac{3}{2}$. If $p$ is the conjugate exponent to $q$ and $\kappa\defin\frac{2}{p}$, then
\begin{equation*}
\frac{1}{p}>1-\frac{2}{3}=\frac{1}{3}, \frac{2-\kappa}{2}q=1, q\sigma>\frac{\sigma+1}{2}
\end{equation*}
H\" older's inequality can now be applied to the last norm in line \ref{lastlinebeforeHolder}.
\begin{align*}
\|(1+(\frac{r_*-\alpha_*}{2M})^2)^{-\sigma} u\|^2=&\int_\starman \frac{|u|^\kappa |u|^{2-\kappa}}{(1+(\frac{\rho_*}{2M})^2)^{2\sigma}}dr_*d^2\omega_{S^2}\\
\leq&\left( \int_\starman |u|^{p\kappa}dr_*d^2\omega_{S^2}\right)^{\frac{1}{p}} \left( \int_\starman \frac{|u|^{(2-\kappa)q}}{(1+(\frac{\rho_*}{2M})^2)^{2\sigma q}}dr_*d^2\omega_{S^2}\right)^\frac{1}{q}\\
\|(1+(\frac{r_*-\alpha_*}{2M})^2)^{-\sigma} u\|\leq& \|u\|^{\frac{1}{p}}\| \frac{|u|}{(1+(\frac{\rho_*}{2M})^2)^{\sigma q}} \|^{1-\frac{1}{p}}\\
\leq& (E^{\frac{1}{2}}T+\|f\|)^\frac{1}{p} \|\frac{u}{(1+(\frac{\rho_*}{2M})^2)^{\frac{\sigma+1}{2}}} \|^{1-\frac{1}{p}}
\end{align*}
For sufficiently large $T$, there is a constant $F$ so that
\begin{equation*}
\|(1+(\frac{r_*-\alpha_*}{2M})^2)^{-\sigma} u\| \leq F T^\frac{1}{p}\|\frac{u}{(1+(\frac{\rho_*}{2M})^2)^{\frac{\sigma+1}{2}}} \|^{1-\frac{1}{p}}
\end{equation*}
\begin{align}
\int_0^T \|(1+(\frac{\rho_*}{2M})^2)^{-\frac{\sigma+1}{2}}u\|^2dt &\leq E^{\frac{1}{2}}(4C_\sigma E^{\frac{1}{2}}+\| f\|) + F T^\frac{1}{p}\|\frac{u}{(1+(\frac{\rho_*}{2M})^2)^{\frac{\sigma+1}{2}}} \|^{1-\frac{1}{p}} \label{BeforeLemma25}
\end{align}

This establishes an integral relation between the local decay norm and its square integral. The local decay norm has bounded derivative since
\begin{equation}
\dt \|\frac{u}{(1+(\frac{\rho_*}{2M})^2)^{\frac{\sigma+1}{2}}} \|^2=2\langle\frac{u}{(1+(\frac{\rho_*}{2M})^2)^{\frac{\sigma+1}{2}}}, \frac{\dot{u}}{(1+(\frac{\rho_*}{2M})^2)^{\frac{\sigma+1}{2}}}\rangle \leq 2 \|\frac{u}{(1+(\frac{\rho_*}{2M})^2)^{\frac{\sigma+1}{2}}} \| E^\frac{1}{2}
\end{equation}
These two conditions are sufficient to apply lemma 25 of \cite{BS}. That lemma states that for $\theta:\Reals\rightarrow\Reals^+$ with uniformly bounded derivative, $\epsilon\in(0,\frac{1}{3})$, if $\int_0^t \theta(\tau)^2d\tau\leq C_1+C_2t^\epsilon\theta^{1-\epsilon}$ then $t^\epsilon\theta(t)^{1-\epsilon}$ goes to zero sequentially and hence $\int_0^t\theta(\tau)^2d\tau\leq C_1$. The lemma can be applied with $\theta$ as the local decay norm, $\frac{1}{p}=\epsilon$, and $C_1$ and $C_2$ as in \ref{BeforeLemma25}. 

This proves the result for $\beta\in(\frac{3}{2},2]$ and for $\rho_*$ instead of $r_*$. Since $(1+(\frac{\rho_*}{2M})^2)^{-\beta}$ is a decreasing function of $\beta$, the result holds for all $\beta>\frac{3}{2}$. Finally since for any $\beta$ there is a constant so that for all $r_*$, $(1+(\frac{r_*-\alpha_*}{2M})^2)^{-\beta}\leq C(1+(\frac{r_*}{2M})^2)^{-\beta}$ the statement of the theorem holds. 
\end{proof}

\appendix
\section{Numerical verification of the positivity of the commutator}

The key step in proving the local decay estimate is the lower bound for the commutator $i[H,\gamma]$ proven in lemma \ref{LMorawetzlike}. From the asymptotics of $i[H_2,\gamma]\like r^{-4}$ and $i[H_3,\gamma]\like r^{-3}$ it is clear that the negative contributions from $i[H_2,\gamma]$ will be dominated eventually and it is sufficient to show $i[H,\gamma]$ is positive in some finite domain. To verify positivity of the commutator, the sum of the exact form for $i[H_2,\gamma]$ from equation \ref{eVlowerbound}, the lower bounds for $i[H_1,\gamma]$ from equation \ref{eH1lowerbound}, and the lower bound for $i[H_3,\gamma]$ from equation \ref{eH3lowerbound} and $l\geq2$ is plotted for $M=1$ and $\sigma=1$. From the graph it is clear that the total commutator is positive. The graph decays because all the terms involved decay. 

\begin{figure}
\begin{center}
\resizebox{6cm}{!}{\includegraphics{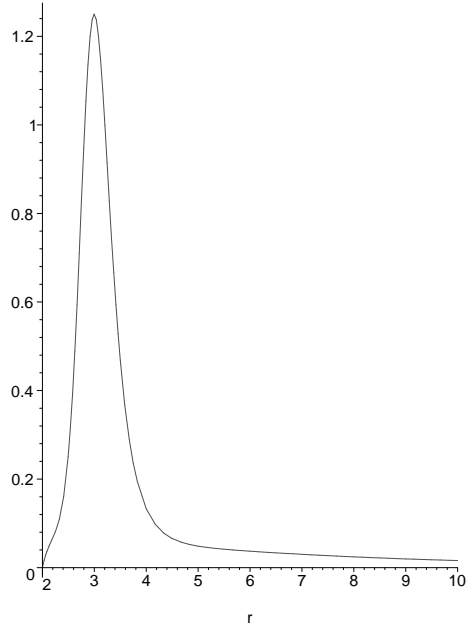}}
\caption{Plot of a lower bound for $i[H,\gamma]$ as a function of $r$. }
\end{center}
\end{figure}

\vspace{.5in}

{\large\bf Acknowledgments:} 
This work was completed when the authors were visiting the Fields Institute in Toronto; we thank the organizers of the special PDE program for their hospitality. We would like to thank the organizers of the Centre de Recherches Mathematiques's workshop on the interaction of gravity with classical fields where this project was started. We thank B.F. Whiting and S. T. Yau for valuable discussions. 

This work was partially supported by NSF grant \# DMS-0100490.

\end{document}